\documentclass[twocolumn,prl,superscriptaddress,showpacs,floatfix,%
preprintnumbers, nofootinbib,hyperref]{revtex4-1} 
\usepackage{graphicx}
\usepackage{float}

\newcommand{\ben}{\begin{equation}}
\newcommand{\een}{\end{equation}}

\newcommand{\be}{\begin{equation}}
\newcommand{\ee}{\end{equation}}  
\newcommand{\bea}{\begin{eqnarray}}
\newcommand{\eea}{\end{eqnarray}}  
\newcommand{\gag}{g_{a\gamma}}

\begin{document}

\title{Enhancing the Spectral Hardening of Cosmic TeV
Photons by Mixing with Axionlike Particles in the
Magnetized Cosmic Web}

\author{Daniele Montanino} \affiliation{Dipartimento di
Matematica e Fisica ``Ennio de Giorgi'', Universit\`a del
Salento, Via Arnesano, I--73100 Lecce, Italy
}
\affiliation{Istituto Nazionale di Fisica Nucleare - Sezione
di Lecce, Via Arnesano, I--73100 Lecce, Italy
}
\author{Franco Vazza} \affiliation{INAF - Istituto di Radio
Astronomia di Bologna, Via Gobetti 101, 40122 Bologna,
Italy} \affiliation{Universit\"{a}t Hamburg, Hamburger
Sternwarte, Gojenbergsweg 112, 20129, Hamburg, Germany}
\author{Alessandro Mirizzi} \affiliation{Dipartimento
Interateneo di Fisica ``Michelangelo Merlin,'' Via Amendola
173, 70126 Bari, Italy.} \affiliation{Istituto Nazionale di
Fisica Nucleare - Sezione di Bari, Via Amendola 173, 70126
Bari, Italy.} \author{Matteo Viel}
\affiliation{SISSA-International School for Advanced
Studies, Via Bonomea 265, 34136 Trieste, Italy}
\affiliation{INAF - Osservatorio Astronomico di Trieste, Via
G. B. Tiepolo 11, I-34143 Trieste, Italy} \affiliation{INFN
- National Institute for Nuclear Physics, via Valerio 2,
I-34127 Trieste, Italy}

\begin{abstract}

Large-scale extragalactic magnetic fields  may induce
conversions between very-high-energy photons and axionlike
particles (ALPs), thereby shielding the photons from
absorption on the extragalactic background light. However,
in simplified ``cell" models, used so far to represent 
extragalactic magnetic fields, this mechanism would be
strongly suppressed  by current astrophysical bounds. Here
we consider a recent model of extragalactic magnetic fields
obtained from large-scale cosmological simulations. Such
simulated magnetic fields would have large enhancement in
the filaments of matter.  As a result, photon-ALP
conversions would produce a significant   spectral hardening
 for cosmic TeV photons. This effect  would be probed with
the upcoming Cherenkov Telescope Array detector. This
possible  detection would give a unique chance to perform a
tomography of  the magnetized cosmic web with ALPs.
\end{abstract}

\maketitle

\emph{Introduction}---Axionlike particles (ALPs) are
ultralight pseudoscalar bosons $a$ with a two-photon vertex
$a \gamma \gamma$, predicted by several extensions of the
Standard Model (see~\cite{Jaeckel:2010ni} for a recent
review). In the presence of an external magnetic field, the
$a \gamma \gamma$ coupling  leads to the phenomenon of
photon-ALP mixing~\cite{Raffelt:1987im}. This effect allows
for the possibility of   direct searches of ALPs in
laboratory  experiments. In this respect a rich, diverse
experimental program is being carried out, exploiting
different sources and
approaches~\cite{Ehret:2010mh,Arik:2015cjv,Duffy:2006aa}.
See~\cite{Graham:2015ouw} for a review.

 Because the $a \gamma \gamma$ coupling, ultralight ALPs can
 also play an important role  in astrophysical observations.
 In particular, an intriguing hint for ALPs has been
 recently suggested by very-high-energy (VHE) $\gamma$-ray
 experiments. In this respect, recent observations of
 cosmologically distant $\gamma$-ray sources by ground-based
 $\gamma$-ray Imaging Atmospheric Cherenkov Telescopes have
 revealed a surprising degree of transparency of the
 Universe to VHE
 photons~\cite{Aharonian:2005gh,Horns:2012fx}, where one
 would have expected a significant absorption of VHE photons
   by pair-production processes ($\gamma_{\rm
 VHE}+\gamma_{\rm EBL} \to e^+ e^-$) on the extragalactic
 background light (EBL). Even though this problem has been
 also analyzed using more conventional physics (see,
 e.g.,~\cite{Essey:2009zg,Essey:2009ju}), photon-ALP
 oscillations  in large-scale magnetic fields provide a
 natural mechanism to drastically reduce  the photon
 absorption~\cite{Csaki:2003ef,De Angelis:2007dy,
 Simet:2007sa,Dominguez:2011xy,DeAngelis:2011id,
 Horns:2012kw,Meyer:2013pny,Troitsky:2015nxa}. In
 order to have efficient conversions one should achieve the
 \emph{strong-mixing} regime which is realized above a
 critical energy given by~\cite{DeAngelis:2007wiw}
\begin{equation}
E_c \simeq \frac{500 \,\ m_a^{2}}{(10^{-9} 
\,\ \textrm{eV})^{2}}\left(\frac{10^{-9} \,\ \textrm{G}}{B_T} \right) 
\left(\frac{5 \times 10^{-11}}{\gag} \right) \,\ \textrm{GeV} \,\ .
\end{equation}
In the previous expression one typically considers $\gag
\lesssim  5 \times 10^{-11}$~GeV$^{-1}$ in order to be
consistent with the bounds from the helioscope experiment
CAST at CERN~\cite{Arik:2015cjv,Castcant},  and from
energy loss in globular-clusters stars~\cite{Ayala:2014pea}
that give $\gag  \lesssim 6 \times 10^{-11}$~GeV$^{-1}$ (see
Fig.~1). Notably two mechanisms have been proposed in order
to reduce the cosmic opacity through conversions into ALPs.
The first involves extragalactic magnetic fields~\cite{De
Angelis:2007dy}.  At present, the lowest upper limits on 
extragalactic magnetic fields on the largest cosmological scales 
come from cosmic microwave background 
observations ~\cite{Planck2015}  and from
Faraday rotation measurements  of polarized extragalactic
sources ~\cite{Blasi:1999hu}  and are compatible with an average
field strength of $B\lesssim {\mathcal O} (1$~nG)  with a
coherence length $l_c \sim {\mathcal O}
(1$~Mpc)~\cite{Pshirkov:2015tua}. In this case, in order to
have sizeable conversions into ALPs at $E\gtrsim 100$~GeV,
one should consider $m_a\lesssim10^{-9} \,\ \textrm{eV}$. 
In the second scenario, one requires strong photon
conversions in the magnetic fields of the source and
back-conversions in the Milky Way~\cite{Simet:2007sa}. 
Estimates on the Galactic magnetic field of order $B\sim {\mathcal
O}(10^{-6}$~G) are obtained by measurements of the Faraday 
Rotation~\cite{Jansson:2012pc,Pshirkov:2011um,
Oppermann:2011td,Mao:2010zr}. In this case,  one expects
significant conversions for $m_a\lesssim10^{-8} \,\
\textrm{eV}$. The resultant parameter space where ALPs would
explain the low $\gamma$-ray opacity under these conditions
is shown in light blue~\cite{Meyer:2013pny} in Fig.~1.

However, the range of the parameters where ALPs would impact
the cosmic transparency  is constrained from other
observations, as shown in Fig.~1. In particular, for ALPs
with masses  $m_a \lesssim10^{-9}$~eV, the strongest bound
on $\gag$ is derived from the absence of $\gamma$-rays from
SN 1987A. In this regard, a recent analysis results in $\gag
< 5 \times 10^{-12}$~GeV$^{-1}$  for  $m_a \lesssim
10^{-10}$~eV~\cite{Payez:2014xsa}. A comparable bound on
$\gag$ has been recently  extended in the mass range $0.5
\lesssim m_a \lesssim 5$~neV from the nonobservation in
Fermi Large Array Telescope (LAT) data of irregularities
induced by photon-ALP conversions in the $\gamma$-ray
spectrum of NGC 1275,  the central galaxy of the Perseus
Cluster~\cite{TheFermi-LAT:2016zue}.
It is worth mentioning that from the absence of x-ray spectral 
modulations in active galactic nuclei,  for $m_a\lesssim10^{-12} 
\,\ \textrm{eV}$ we have a stronger bound on the coupling:
$\gag \lesssim 1 \times 10^{-12}$~GeV$^{-1}$~\cite{Berg:2016ese,
Conlon:2017qcw}. However,
in the following we will always refer to higher ALP masses.
Data from the H.E.S.S. observations of the distant BL Lac
object PKS 2155-304 also limit $\gag < 2.1 \times
10^{-11}$~GeV$^{-1}$ for  $15 \lesssim m_a \lesssim
60$~neV~\cite{Abramowski:2013oea}. Therefore, while the
mechanism of conversions into the Galactic magnetic field 
is still valid at $m_a\lesssim10^{-8} \,\ \textrm{eV}$,  
the conversions into the extragalactic magnetic field seem
strongly disfavored as a mechanism to reduce the cosmic
opacity. However, the distribution of the extragalactic
magnetic fields has been oversimplified in previous works on
ALP conversions. In particular, a cell-like structure
(hereafter  named the ``cell" model) has been adopted with 
many domains of equal size ($l_c\sim 1$~Mpc) in which the
magnetic field has (constant) random values and
directions~\cite{De Angelis:2007dy}. Only recently it has
been pointed out  that in more realistic situations, the
magnetic field direction would  vary continuously along the
propagation path, and this would lead to sizeable
differences in the ALP conversions with respect to the
``cell"
model~\cite{Angus:2013sua,Wang:2015dil,Kartavtsev:2016doq,
Masaki:2017aea}. Here we study for the first time the photon
conversions into ALPs using recent magneto-hydrodynamical
cosmological simulations~\cite{Vazza14, Gheller16}, which represent a
step forward in the realistic modeling of cosmic magnetism.
In this more realistic case the magnetic fields can locally
fluctuate in filaments of matter up to 2 orders of
magnitude larger that found in the ``cell" model and
photon-ALP conversions are enhanced compared to previous
estimates.  Indeed, significant conversions are found both
in the low-mass region (indicated as ``LM,'' $m_a \lesssim
10^{-10}$~eV) below the SN 1987A bound and in the high-mass
region (``HM," $m_a \gtrsim 10^{-8}$~eV)  on the right of 
the recent Fermi-LAT bound. These two ranges are indicated
with small squares  in Fig.~1 and represent regions of
sensitivity (see Supplemental Material for details). We
have also checked that for these parameters possible
photon-ALP conversions at relatively high redshifts do not
distort the cosmic microwave background
\cite{Mirizzi:2005ng}. Therefore, our new model
significantly enlarges the parameter space where the
photon-ALP conversions would reduce the cosmic opacity, 
as shown in Fig.~1 where we compare the photon transfer function 
$T_\gamma$ in presence of ALP conversions for the cell model and for 
the simulated model of extragalactic $B$-field (see Supplemental Material for details).
In particular, these values of ALPs parameters are in the reach
of  the planned upgrade of the photon regeneration
experiment ALPS at DESY~\cite{Ehret:2010mh} and  with  the
next generation solar axion detector IAXO (International
Axion Observatory)~\cite{Irastorza:2011gs} (see Fig.~1) or
with Fermi-LAT if  luckily a galactic supernova would
explode in its field of view~\cite{Meyer:2016wrm}. 

\begin{figure}[tbp]
\centering
\includegraphics[angle=0,width=1.\columnwidth]{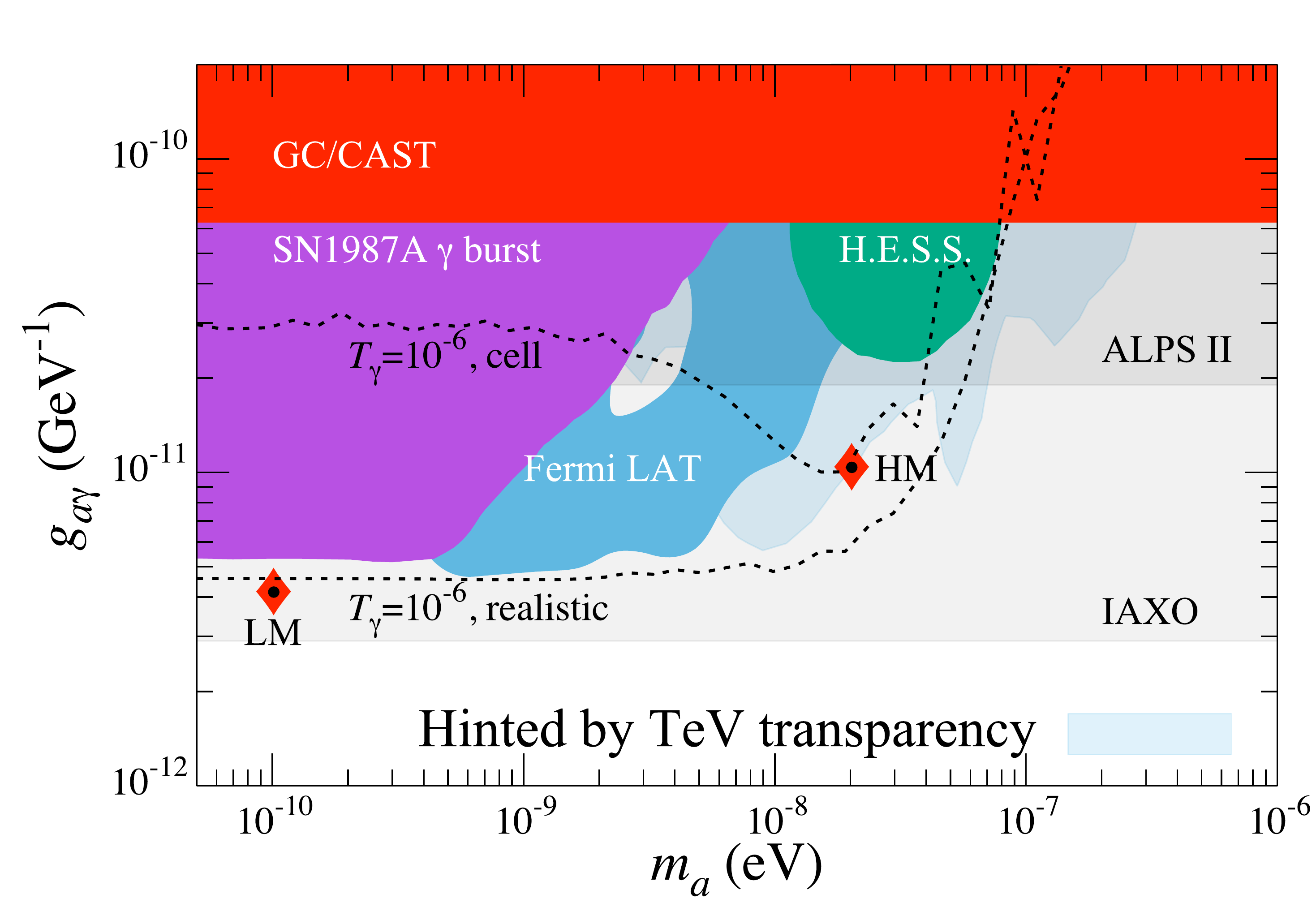}
\vspace{-0.3cm}
\caption{Limits on ALP parameter space in the plane $(m_a,
\gag)$. The parameter space where ALPs could explain the low
$\gamma$-ray opacity is shown in light
blue~\cite{Meyer:2013pny}. The horizontal grey bands
represent the sensitivity of future ALPS-II and IAXO
experiments.
Dashed lines represent the photon transfer function $T_\gamma$ for the cell model 
and simulated model of the extragalactic
$B$-field. The two small squares represent cases of
low-mass and high-mass ALPs where conversions in
simulated extragalactic magnetic fields  would affect the
TeV photon transparency. (See text for details.)}
\label{fig:Excl}
\vspace{-0.3cm}
\end{figure}

\emph{Simulated extragalactic magnetic fields}---Our recent
model for extragalactic magnetic fields comes from a suite of
simulations \cite{Vazza14} produced using the cosmological
code {\large E}{\small NZO} \cite{Enzo14}, with the
magnetohydrodynamical method outlined in \cite{Wang09}.
These simulations are nonradiative and evolved a uniform
primordial seed field of $B_0=1 ~\rm nG$ (comoving) for each
component, starting from $z=38$ and for a comoving volume of
$200^3 \rm Mpc^3$, sampled using a fixed grid of $2400^3$
cells (for the fixed comoving resolution of $83.3$ kpc/cell)
and $2400^3$ dark matter particles. We consider this model
more realistic than other models used in the literature of
ALPS studied, because it incorporates the dynamical
interplay between structure formation (e.g. gas compression
onto filaments, galaxy groups and clusters), rarefactions
(onto voids) and further dynamo amplification where
turbulence is well resolved (typically limited to the most
massive clusters in the volume). More details on the
simulation are given in the Supplemental Material.

\begin{figure}[t]
\centering
\includegraphics[angle=0,width=1.\columnwidth]{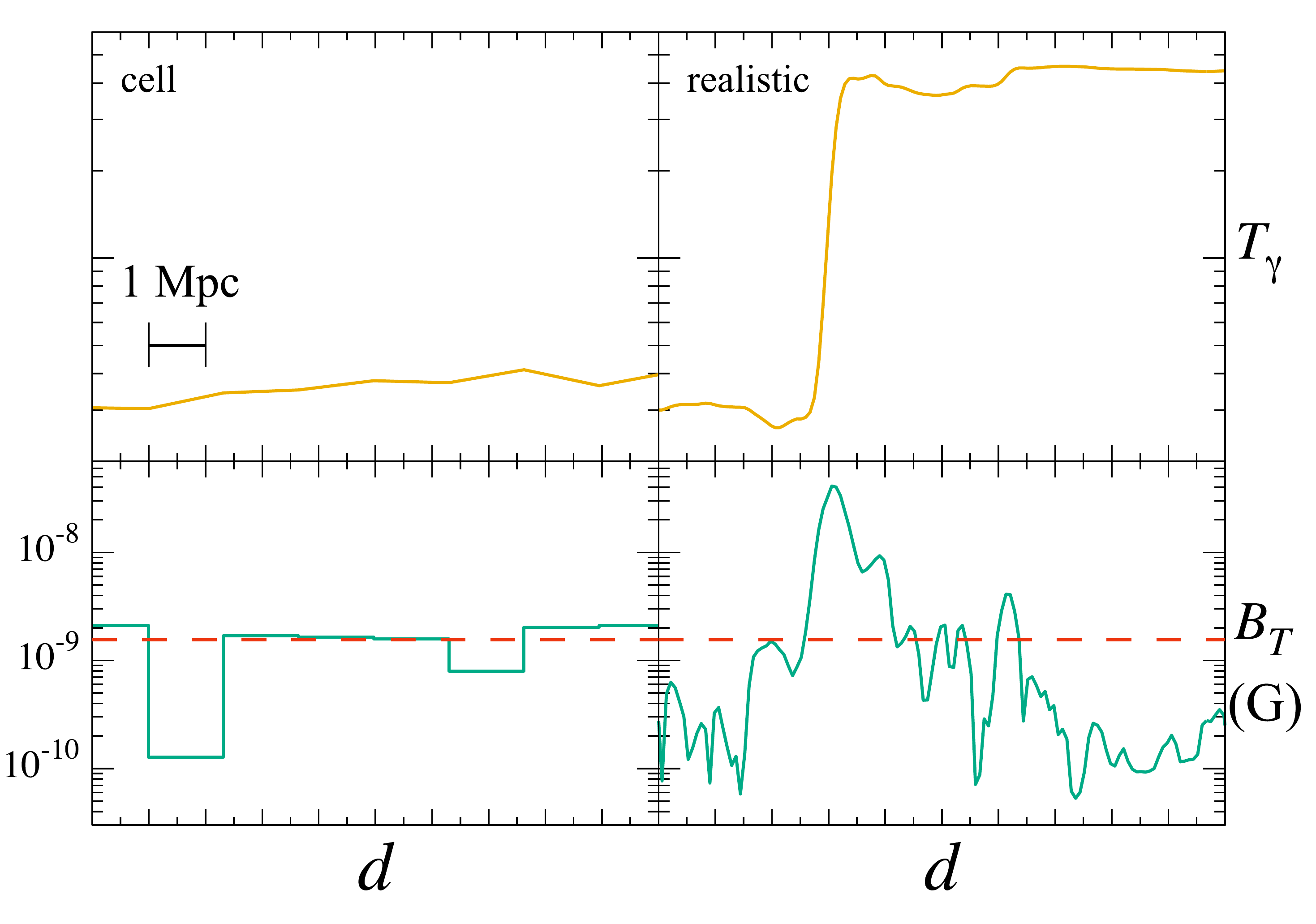}
\vspace{-0.6cm}
\caption{Photon transfer function $T_\gamma$ in function of
the distance $d$  for a given line of sight (upper panels)
and  corresponding variations of the extragalactic magnetic
field (lower panels). We have taken the photon energy
$E=10$~TeV and ALP parameters $m_a = 2 \times 10^{-8}$~eV
and $\gag = 10^{-11}$~GeV$^{-1}$ (HM).   Left panels refer
to the ``cell" model, while right panels represent the more
realistic model of our cosmological simulation. In both the
cases we have chosen as average value of the magnetic field
$\langle B_T \rangle =1.6\times 10^{-9}$~G. }
\label{fig:fig1}
\vspace{-0.3cm}
\end{figure}

\emph{Results}---We closely follow the technique described
in \cite{Horns:2012kw,Mirizzi:2009aj}  to solve the 
propagation equations for the photon-ALP ensemble. We
document this procedure in the Supplemental Material. In
order  to account the VHE photon absorption  we employ the
EBL model of Ref.~\cite{Franceschini:2008tp} as our
benchmark. With respect to 
\cite{Horns:2012kw,Mirizzi:2009aj} we also include the
refraction effect on VHE  photons, recently calculated
in~\cite{Dobrynina:2014qba}. This term would be dominated by
the energy density of  cosmic microwave background photons.
It is expected to damp the photon-ALP conversions at
$E\gtrsim 1$~TeV~\cite{Kartavtsev:2016doq}.
In Fig.~2 we show the photon transfer function $T_\gamma$ in
a certain range $d$ of its path along  a given line of sight
(upper panels) in a region in which the extragalactic
magnetic field has significant variations (lower panels). We
have taken the photon energy $E=10$~TeV. We have chosen as
ALP parameters $m_a = 2 \times 10^{-8}$~eV and $\gag =
10^{-11}$~GeV$^{-1}$, corresponding to the high-mass (HM)
square in  Fig.~1.   Left panels refer to the ``cell" model,
while right panels represent the model from the cosmological
run. The ``cell" model has been generated using a fixed 
(comoving) size of $l_c=1.4$~Mpc per cell, which is the
typical coherence length of magnetic fields in the newly
simulated model, based on spectral analysis. The transverse
magnetic field in each cell is chosen as
$B_T=B_{0}\sin\theta$ where $\theta$ is a random zenithal
angle in $[0,\pi)$ and $B_{0}=1.9\times 10^{-9}$~G is the
rms of the magnetic field of all realistic realizations.
The  dashed line in Fig.~2 corresponds to $\langle B_T
\rangle=\sqrt{2/3}\, B_{0} =1.6\times 10^{-9}$~G which is
thus in both cases the ensemble average of the  strength
of the transverse magnetic field. It can be seen that  while
in the ``cell" model the variations of the $B$ field
saturate the maximum value ($B_0$), in the more realistic
scenario generated by the simulations there are several
peaks where $B_T \sim 10^{-7}$~G, which are found to
correlate with gas structures in the cosmic web (i.e.\
mostly filaments and outer regions of galaxy clusters and
groups). As a consequence, while in the ``cell" model
$T_\gamma$ does not show sizeable variations since the
strong mixing regime is suppressed by $m_a$ [see Eq.~(1)],
in the more realistic case  one finds a significant jump in
$T_\gamma$ when the peak in the magnetic field is reached. 
Indeed, since the conversion probability in ALPs scales as
$P_{a\gamma}  \sim (\gag B_T d)^2 $, when the strong peak in
$B_T$ is encountered a strong photon regeneration takes
place.

\begin{figure}[b]
\centering
\includegraphics[angle=0,width=1.\columnwidth]{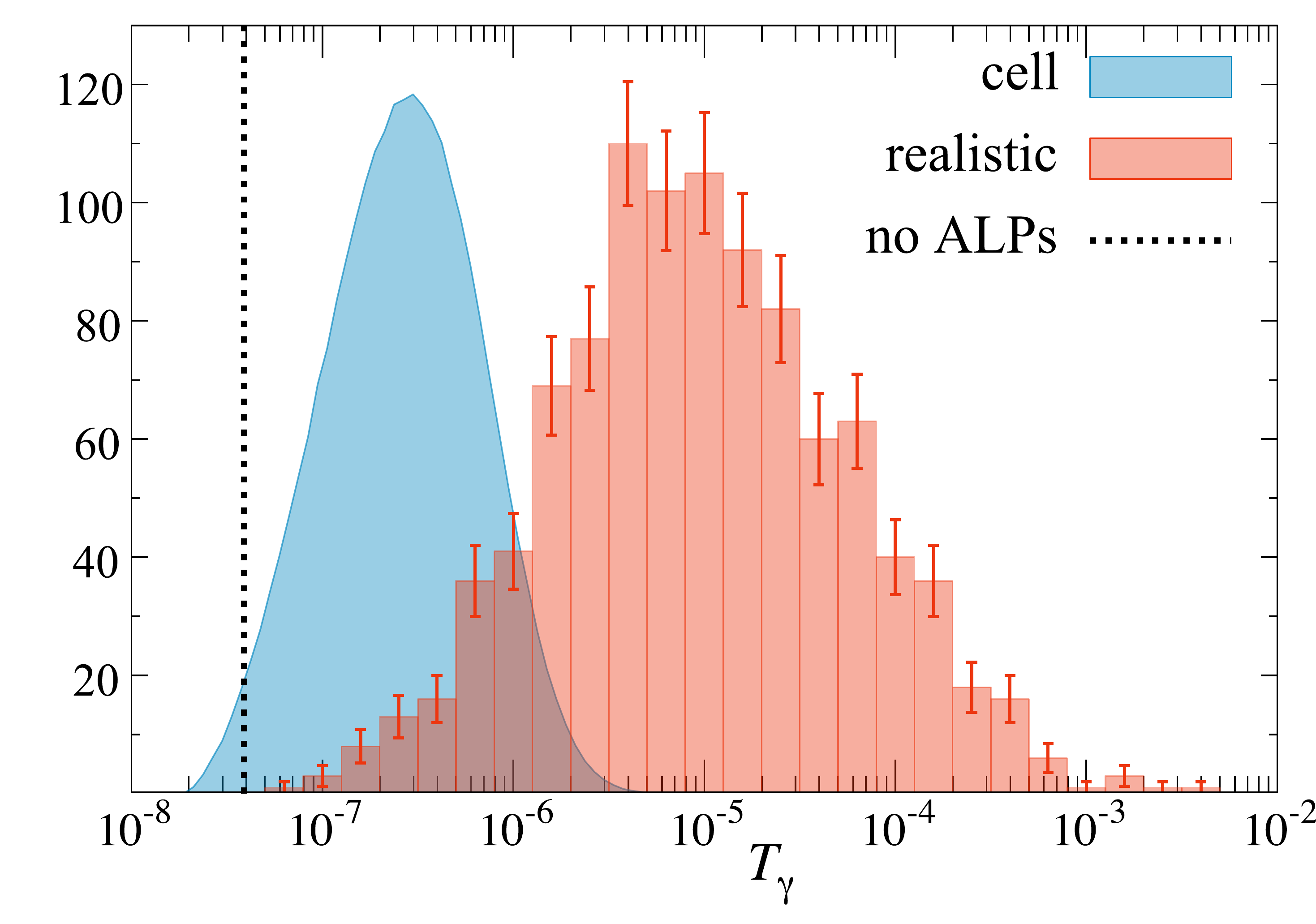}
\vspace{-0.6cm}
\caption{Probability distribution functions  for the transfer function $T_\gamma$ for
photons emitted from a source at redshift $z=0.3$, with
$E=10$~TeV and ALPs parameters $m_a = 2 \times 10^{-8}$~eV
and $\gag = 10^{-11}$~GeV$^{-1}$ (HM). Left distribution
refers to the cell model, while the right one to the more
realistic case. The vertical line corresponds to the no ALPs
case.}
\label{fig:fig2}
\end{figure}

\begin{figure*}[tbp]
\centering
\includegraphics[angle=0,width=1.7\columnwidth]{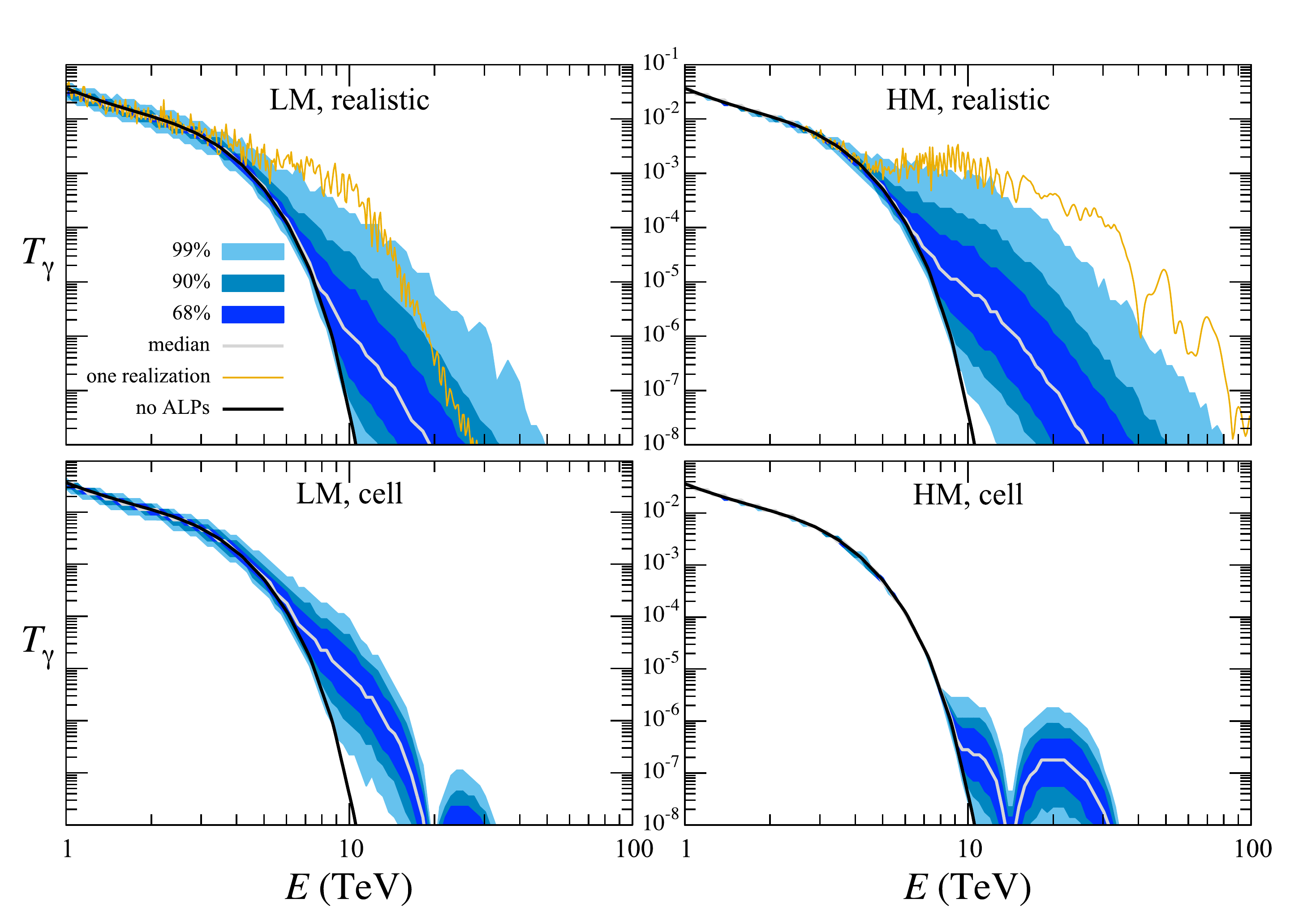}
\vspace{-0.3cm}
\caption{Photon transfer function $T_\gamma$ as a function
of photon energy for a source at redshift $z=0.3$ for $m_a =
 10^{-10}$~eV and $\gag = 4 \times 10^{-12}$~GeV$^{-1}$ (LM,
left panels) and  $m_a = 2 \times 10^{-8}$~eV and $\gag =
10^{-11}$~GeV$^{-1}$ (HM, right panels). The upper panels refer
to more realistic models of an extragalactic magnetic field,
while lower panels are for the cell model. The black
curve corresponds to the case of only  absorption onto EBL.
The solid grey curve represents the median $T_\gamma$ in the
presence of ALPs conversions. The orange curve corresponds
to conversions for a particular realization of the
extragalactic magnetic field. The shaded band is the
envelope of the results on all the possible realizations of
the extragalactic magnetic field at 68 \% (dark blue), 90 \%
(blue)  and 99 \% (light blue) C.L., respectively.} \label{fig:fig3}
\end{figure*}

Because of the random orientation of the magnetic field, the
effect of photon-ALP conversions strongly depends on the
orientation of the line of sight. Therefore the photon flux
observed at Earth should be better characterized in terms of
probability distribution functions, obtained by considering
$\gamma \to a$ conversions over different realizations of
the extragalactic magnetic field. An example of these
distributions is shown in Fig.~3 for a source at redshift
$z=0.3$ and energy $E=10$~TeV. We have fixed ALP parameters
to the HM case. In the ``cell'' case, the PDF has been
found by simulating $10^6$  different realizations of the
extragalactic magnetic field. Conversely, in the realistic case,
the PDF has been obtained 
extracting $10^3$ onedimensional beams of
cells randomly extracted from the outputs of the
cosmological simulation at increasing redshift. The $y$ axis shows
the number counting for the realistic case only.
The vertical thick line corresponds to the no ALPs
case. We see that ALP conversions in dynamical magnetic
field and in the cell model produce probability
distributions which are strongly different. In particular
for the cell model the peak in the distribution is
$T_\gamma\simeq 3\times 10^{-7}$  with a 90 \% C.L.\ in the
interval $5.6\times 10^{-8}$--$1.1\times 10^{-6}$, while in
the more realistic case one finds the peak at $T_\gamma
\sim3\times 10^{-5}$  and a 90 \% C.L.\ in the interval
$4.5\times 10^{-7}$--$1.4\times 10^{-4}$. It is evident that a
strong enhancement of the transfer function in the more
realistic case would imply a significant hardening of
the photon spectra with respect to the cell model.
Note that in the realistic case the statistical error in each bin is due to the
limited number of realizations of the magnetic fields
configurations.

In order to discuss the observational signatures in the
energy spectra of VHE photon sources we present  in Fig.~4
the distribution of photon transfer function $T_\gamma$  in
function of the photon energy for a source at redshift
$z=0.3$ obtained with $10^3$ realizations. In the left
panels we consider $m_a =  10^{-10}$~eV and $\gag = 4 \times
10^{-12}$~GeV$^{-1}$, corresponding to the LM case of Fig.~1,
while in the right panels we take  HM parameters. Upper
panels refer to our simulations for magnetic field, while
lower ones are for the cell model. The black solid curve
represents the $T_\gamma$ expected in the presence of only
absorption onto EBL. The solid grey curve represents the
median $T_\gamma$ in the presence of ALPs conversions. The
orange curve corresponds to conversions for a particular
realization of the extragalactic magnetic field. The shaded
band is the envelope of the results on all the possible
realizations of the extragalactic magnetic field at 68 \%
(dark blue), 90 \% (blue) and 99 \% (light blue) C.L.,
respectively. According to conventional physics, it turns
out that the  $T_\gamma$ gets dramatically suppressed at
high energies ($E \gtrsim 4$ TeV). As expected, including
ALP conversions with  cell  magnetic fields, the enhancement
of $T_\gamma$ with respect to the standard case is modest
since it is suppressed by the small coupling (left panel) or
the high ALP mass (right panel). However, when we consider
ALP conversions in more realistic magnetic fields  the
enhancement of $T_\gamma$ is striking. ALP conversions in
such models would produce a considerable hardening of the
spectrum at high enough energies, thereby making it possible
to detect VHE photons in a range where no observable signal
would be expected according to conventional physics or to
conversions with cell magnetic fields. An example of a
particular realization is shown by the orange curve. In this
specific case we see that the observable photon flux at high
energies can be significantly larger than the average one.
On this specific line of sight the enhancement of  
$T_\gamma$ with respect to the standard case would  reach 3
orders of magnitude. Note also the rapid oscillations
observed in the  $T_\gamma$. These are induced by the the
photon dispersion effect on cosmic microwave background
\cite{Kartavtsev:2016doq} (see also Supplemental Material)
and would leave observable signatures on the VHE photon
spectra unexpected in the standard case. Depending on the
particular magnetic realization crossed by the photons, it
is also possible to observe a suppression of the photon flux
stronger than in the presence of conventional physics.
Nevertheless,  from Fig.~4 one infers that the cases in
which $T_\gamma$ is enhanced at high energies are much more
probable. 


\emph{Conclusions}---We have studied the conversions of VHE
photons into  ALPs  proposed as a mechanism to reduce the
absorption onto EBL, using for the first time more realistic
models of extragalactic magnetic fields, obtained from the
largest magnetohydrodynamical cosmological simulations in
the literature. We find an enhancement of the magnetic field
with respect to what was predicted in the naive cell model, 
due to the fact that simulated magnetic fields display
larger fluctuations, correlated with density fluctuations of
the cosmic web. This effect would give a significant boost
to photon-ALP conversions.
Indeed, using more realistic models of the magnetic field we
have found significant conversions also in regions of the
parameter space consistent with previous astrophysical 
bounds. This mechanism would produce a significant hardening
of the VHE photon spectrum from faraway sources, and we
expect such signature to emerge at energies $E\gtrsim
1$~TeV. Therefore, this scenario is testable with the
present generation of the Imaging Atmospheric Cherenkov
Telescope, covering energies in the range from $\sim$~50~GeV
to $\sim$~50~TeV \cite{Meyer:2014gta,Consortium:2010bc}.

\emph{Acknowledgments}---We thank Maurizio Giannotti,
Manuel Meyer, Georg Raffelt and Pasquale Dario Serpico for
reading our manuscript and for useful comments on it. We
acknowledge the important collaboration of Claudio Gheller
and Marcus Br\"{u}ggen in the production of the cosmological
simulation analyzed in this work. The work of A.M.\ and
D.M.\ is supported by the Italian Ministero dell'Istruzione,
Universit\`a e Ricerca (MIUR) and Istituto Nazionale di
Fisica Nucleare (INFN) through the ``Theoretical
Astroparticle Physics'' project. F.V.\ acknowledges the use
of computing resources on Piz Daint (ETHZ-CSCS, Lugano)
under allocation s585 and s701. F.V.\ acknowledges financial
support from the grant VA 876-3/1 by DFG, and from the
European Research Council (ERC) under the
European Union's Horizon 2020 research and innovation
program under the Marie-Sklodowska-Curie Grant Agreement
No.~664931 and under the ``MAGCOW" 
Starting Grant No.~714196. M.V. is supported by PRIN INAF, PRIN MIUR,
INDARK-PD51 and ERC "cosmoIGM" grants.

\section*{Supplemental Material}

\emph{Setup of photon-ALP oscillations}---Photon-ALP mixing 
occurs in the presence of an external
magnetic field ${\bf B}$ due to the interaction
term~\cite{Raffelt:1987im}
\begin{equation}
\label{mr}
{\cal L}_{a\gamma}=-\frac{1}{4} \,\gag
F_{\mu\nu}\tilde{F}^{\mu\nu}a=\gag \, {\bf E}\cdot{\bf B}\,a~,
\end{equation}
where $\gag$ is the photon-ALP coupling constant (which has
the dimension of an inverse energy).

We consider throughout a monochromatic photon/ALP beam of
energy $E$ propagating along the $x_3$ direction in a cold
ionized and magnetized medium. It has been shown that for
very relativistic ALPs and {\it polarized} photons, the beam
propagation equation can be written in a Schr\"odinger-like
form in which $x_3$ takes the role of
time~\cite{Raffelt:1987im,Kartavtsev:2016doq} 
\begin{equation}
\label{we} i  \frac{d}{d x_3}  \left(\begin{array}{c}A_{1}
(x_3) \\ A_2 (x_3) \\ a (x_3) \end{array}\right)  = \left(
{\cal H}_{\rm disp} - \frac{i}{2}  {\cal H}_{\rm abs} 
\right)  \left(\begin{array}{c}A_{1} (x_3) \\ A_2 (x_3) \\ a
(x_3) \end{array}\right) ~,
\end{equation}
where $A_1(x_3)$ and $A_2 (x_3)$ are the  photon linear
polarization amplitudes along the $x_1$ and $x_2$ axis,
respectively, $a (x_3)$ denotes the ALP amplitude. The
Hamiltonian $ {\cal H}_{\rm disp} $ represents the
photon-ALP dispersion matrix, including the mixing  and the
refractive effects,   while ${\cal H}_{\rm abs}$  accounts
for the photon absorption effects on the low-energy photon
backgrounds. We denote by $T (x_3,0;E)$ the transfer
function, namely the solution of Eq.~(\ref{we}) with initial
condition $T (0,0;E) = 1$. 

The Hamiltonian $ {\cal H}_{\rm disp} $ simplifies if we
restrict our attention to the case in which ${\bf B}$ is
homogeneous. We denote by ${\bf B}_T$ the transverse
magnetic field, namely its component in the plane normal to
the beam direction and we choose the $x_2$-axis along ${\bf
B}_T$ so that $B_1$ vanishes. The linear photon polarization
state parallel to the transverse field direction ${\bf B}_T$
is then denoted by $A_{\parallel}$ and the orthogonal one by
$A_{\perp}$. Correspondingly, the mixing matrix can be
written as~\cite{Raffelt:1987im,Kartavtsev:2016doq} 
\begin{equation}
{\cal H}_{\rm disp} =   \left(\begin{array}{ccc}
\Delta_{ \perp}  & 0 & 0 \\
0 &  \Delta_{ \parallel}  & \Delta_{a \gamma}  \\
0 & \Delta_{a \gamma} & \Delta_a 
\end{array}\right)~,
\label{eq:massgen}
\end{equation}
whose elements are~\cite{Raffelt:1987im} $\Delta_\perp
\equiv \Delta_{\rm pl} + \Delta_{\perp}^{\rm CM} +
\Delta_{\rm CMB},$ $ \Delta_\parallel \equiv \Delta_{\rm pl}
+ \Delta_{\parallel}^{\rm CM} + \Delta_{\rm CMB},$
$\Delta_{a\gamma} \equiv {g_{a\gamma} B_T}/{2} $ and
$\Delta_a \equiv - {m_a^2}/{2E}$, where $m_a$ is the ALP
mass. The term $\Delta_{\rm pl} \equiv -{\omega^2_{\rm
pl}}/{2E}$ accounts for plasma effects, where $\omega_{\rm
pl}$ is the plasma frequency expressed as a function of the
electron density in the medium $n_e$ as $\omega_{\rm pl}
\simeq 3.69 \times 10^{- 11} \sqrt{n_e /{\rm cm}^{- 3}} \,
{\rm eV}$. The  terms $\Delta_{\parallel,\perp}^{\rm CM}$
describe the Cotton-Mouton  effect, i.e.~the birefringence
of fluids in the presence of a transverse magnetic field.  A
vacuum Cotton-Mouton effect is expected from QED one-loop
corrections to the photon polarization in the presence of an
external magnetic field $\Delta_\mathrm{QED} =
|\Delta_{\perp}^{\rm CM}- \Delta_{\parallel}^{\rm CM}|
\propto B^2_T$, but this effect is completely negligible in
the present context. Recently it has been realized that also
background photons can contribute to the photon
polarization. At this regard a guaranteed contribution is
provided by the CMB radiation, leading to $\Delta_{\rm CMB}
\propto \rho_{\rm CMB}$ \cite{Dobrynina:2014qba}. We will
show how this term would play a crucial role for the
development of the conversions at  high energies. An
off-diagonal $\Delta_{R}$ would induce the Faraday rotation,
which is however totally irrelevant at VHE, and so it has
been dropped. For our benchmark values corresponding to the
HM point, numerically we find 
\begin{eqnarray}  
\Delta_{a\gamma}&\simeq &   1.5\times10^{-2} 
\left(\frac{g_{a\gamma}}{10^{-11}\textrm{GeV}^{-1}} \right)
\left(\frac{B_T}{10^{-9}\,\rm G}\right) {\rm Mpc}^{-1}
\nonumber\,,\\  
\Delta_a &\simeq &
 -3.2 \times 10^{1} \left(\frac{m_a}{2 \times 10^{-8} 
        {\rm eV}}\right)^2 \left(\frac{E}{{\rm TeV}} \right)^{-1} 
        {\rm Mpc}^{-1}
\nonumber\,,\\  
\Delta_{\rm pl}&\simeq & 
  -1.1\times10^{-7}\left(\frac{E}{{\rm TeV}}\right)^{-1}
         \left(\frac{n_e}{10^{-3} \,{\rm cm}^{-3}}\right) 
         {\rm Mpc}^{-1}
\nonumber\,,\\
\Delta_{\rm QED}&\simeq & 
4.1\times10^{-9}\left(\frac{E}{{\rm TeV}}\right)
\left(\frac{B_T}{10^{-9}\,\rm G}\right)^2 {\rm Mpc}^{-1} \nonumber\,,\\
\Delta_{\rm{CMB}}&\simeq & 0.80 \times 10^{-1} \left(\frac{E}{{\rm TeV}} 
\right)  {\rm Mpc}^{-1} \,\ .
\label{eq:Delta0}\end{eqnarray}

VHE photons undergo  pair production absorption by EBL low
energy photons $\gamma_{\rm VHE}\gamma_{\rm EBL}\to e^+e^-$,
 dominated by the interactions with optical/infrared EBL
photons. The absorptive part of the Hamiltonian can be
written in the form
\begin{equation}
{\cal H}_{\rm abs} =   \left(\begin{array}{ccc}
\Gamma  & 0 & 0 \\
0 &  \Gamma  & 0  \\
0 & 0 & 0 
\end{array}\right)~,
\label{eq:abs}
\end{equation}
where $\Gamma$ is the photon absorption rate
(see~\cite{Mirizzi:2009aj} for details). Several realistic
models for the EBL are available in the literature, which
are basically in mutual agreement. Among all possible
choices, we employ the EBL model~\cite{Franceschini:2008tp}
as our benchmark. For crude numerical estimates at zero
redshift we use for the absorption rate~\cite{Mirizzi:2009aj}
\begin{equation}
\Gamma = 1.1 \times 10^{-3} \left(\frac{E}{\rm TeV}\right)^{1.55}
{\rm Mpc}^{-1}  \,\ .
\end{equation}

\emph{Single magnetic domain}---Considering the propagation 
of photons in a single magnetic
domain with a uniform ${\bf B}$-field with $B_1 = 0$, the 
component $A_{\perp}$ decouples away, and the propagation
equations reduce to a 2-dimensional problem. Its solution
follows from the diagonalization of the Hamiltonian through
a similarity transformation performed with an orthogonal
matrix, parametrized by the (complex) rotation angle
$\Theta$ which takes the
value~\cite{Raffelt:1987im,Kartavtsev:2016doq}
\begin{equation}
\Theta = \frac{1}{2}\textrm{arctan}\left(\frac{2 \Delta_{a
\gamma}}{\Delta_{\parallel}-\Delta_{a} - \frac{i}{2}
\Gamma}\right) \,\ . \label{theta}
\end{equation}
Note that  $\Delta_a  < 0$ and $\Delta_{\parallel} >0$.
Therefore, these two contributions always sum and   must be
separately small to achieve large mixing angle. When $
\Delta_{a \gamma} \gg \Delta_{\parallel}-\Delta_{a}$ the
photon-ALP mixing is close to maximal, $\Theta \to \pi/4$
(if the absorption is small as well). On the other hand,
from Eq.~(\ref{eq:Delta0}) one sees that $\Delta_{\rm CMB}$
grows   linearly with the photon energy. Therefore  at
sufficiently high energies $ \Delta_{a \gamma} \ll
\Delta_{\parallel}-\Delta_{a}$  and the photon-ALP mixing is
suppressed.

One can introduce a generalized (including absorption) 
photon-ALP oscillations frequency
\begin{equation}
\Delta_{\rm osc} \equiv \left[(\Delta_{\parallel}
-\Delta_{a} -\frac{i}{2} \Gamma)^2 + 4 \Delta_{a \gamma}^2
\right]^{1/2}~. \label{eq:deltaosc}
\end{equation}

In particular, if absorption effect are small the
probability for a photon emitted in the state
$A_{\parallel}$ to oscillate into an ALP after traveling a
distance $d$ is given by~\cite{Raffelt:1987im}
\begin{eqnarray}
P_{\gamma \to a}^{(0)}  &=& {\rm sin}^2 2 \Theta \  {\rm
sin}^2 \left( \frac{\Delta_{\rm osc} \, d}{2} \right)  \,\
\nonumber \\ &=& (\Delta_{a \gamma} d)^2
\frac{\sin^2(\Delta_{\rm osc} d/2)}{(\Delta_{\rm osc}
d/2)^2} \,\ , \label{conv}
\end{eqnarray}
where in  the oscillation wave number and mixing angle we set $\Gamma=0$.

From  Eq.~(\ref{eq:Delta0})   one would realize that   for
$E\gtrsim 10$~TeV and $B_T \sim 10^{-7}$~G, 
$\Delta_{a\gamma} \gg \Delta_a, \Delta_{\rm pl}$. Therefore,
neglecting the $\Delta_{\rm CMB}$ refractive term one would
obtain $P_{\gamma \to a}^{(0)} \simeq (\Delta_{a \gamma}
d)^2$, that is energy-independent. However, we see that 
$\Delta_{\rm CMB}$ is not negligible at these energies and
would produce peculiar energy-dependent oscillations
imprinting significant features in the VHE photon spectra.

So far, we have been dealing with a beam containing
polarized photons, but since at VHE the polarization cannot
be measured we better assume that the beam is unpolarized.
This is properly done by means of the polarization density
matrix 
\begin{equation}
\rho (x_3) = \left(\begin{array}{c}A_1 (x_3) \\ A_2 (x_3) \\
a (x_3) \end{array}\right) \otimes \left(\begin{array}{c}A_1
(x_3)\  A_2 (x_3)\ a (x_3)\end{array}\right)^{*}
\end{equation}
which obeys the Liouville equation~\cite{Kartavtsev:2016doq}
\begin{equation}
\label{vne} i \frac{d \rho}{d x_3} = [{\cal H}_{\rm disp},
\rho] - \frac{i}{2} \{{\cal H}_{\rm abs}, \rho\}
\end{equation}
associated with Eq. (\ref{we}). Then it follows that the
solution of Eq. (\ref{vne}) is given by 
\begin{equation}
\label{t5L}
\rho (x_3,E) = T (x_3, 0;E) \, \rho (0) \, T^{\dagger}(x_3, 0;E)~, 
\end{equation}
where $\rho (0)$ is the initial beam state. Note that for a
uniform ${\bf B}$ even if we clearly have
\begin{equation}
\label{t5Lq} T (x_3, 0;E) = e^{-i({\cal H}_{\rm disp} -
\frac{i}{2} {\cal H}_{\rm abs}) x_3}~.
\end{equation}

\emph{Magnetized cosmic web}---In the problem discussed 
in this paper we consider
oscillations of  VHE photons into ALPs in the extragalactic
magnetic fields. Therefore we have to deal with a more
general situation than the one depicted in the previous
Section. Indeed, as discussed, the extragalactic magnetic
field is not constant along the photon line of sight. In the
``cell" model  the magnetic field can be modeled as a
network of  a magnetic domains with size set by its
coherence length. Although  $|{\bf B}|\equiv B_0$ is
supposed to be the same in every domain, its direction
changes randomly from one domain to another. Therefore the
propagation over many magnetic domains is clearly a truly
3-dimensional problem, because -- due to the randomness of
the direction of ${\bf B}$ --the  same photon polarization
states play the role of either $A_{\parallel}$ and 
$A_{\perp}$ in different domains. Therefore the Hamiltonian
${\cal H}_{\rm disp}$ entering propagation equation cannot
be reduced to a block-diagonal form similar to
Eq.~(\ref{eq:massgen}) in all domains. Rather, we take the
$x_1$, $x_2$, $x_3$ coordinate system as fixed once and for
all, and -- denoting ${\bf b}_k$ a random unit vector inside
each cell, during their path with a total length $L$ along
the line of sight, the beam crosses $n = L/l_c$ domains,
where $l_c$ is the size of each domain: The set $ \{{\bf
B}_k \}_{1 \leq k \leq n}=\{B_0\, {\bf b}_k \}_{1 \leq k
\leq n}$ represents a given random realization of the beam
propagation. Accordingly, in each domain the Hamiltonian 
${\cal H}_{\rm disp}$  takes the form~\cite{Mirizzi:2007hr}
\begin{equation}
\label{aa8MR}
{\cal H}_{\rm disp}^k = \left(
\begin{array}{ccc}
\Delta_{xx} & \Delta_{xy} & \Delta_{a\gamma} \, \sin\phi_k\\
\Delta_{yx} & \Delta_{yy} & \Delta_{a\gamma} \, \cos\phi_k\\
\Delta_{a\gamma} \, \sin\phi_k& \Delta_{a\gamma} \, 
\cos\phi_k& \Delta_{a} \\
\end{array}
\right)~,
\end{equation} 
where $\phi_k\in[0,2\pi)$ is the  azimuthal (random) angle
between the projection of ${\bf b}_k$ on the $(x_1,x_2)$
plane and the $x_2$ axis, and
\begin{equation}
\Delta_{xx} = \Delta_\parallel \, \sin^2 \phi_k + 
\Delta_\perp \cos^2 \phi_k~,
\end{equation}
\begin{equation}
\Delta_{xy} = \Delta_{yx}=(\Delta_\parallel -\Delta_\perp) 
\sin\phi_k \, \cos\phi_k~,
\end{equation}
\begin{equation}
\Delta_{yy} = \Delta_\parallel \cos^2 \phi_k + 
\Delta_\perp \sin^2 \phi_k~,
\end{equation}
while $\Delta_{a\gamma}$ is given by the first of
Eqs.~(\ref{eq:Delta0}) with $B_T=B_0\sin\theta_k$, where
$\theta_k$ is the zenith angle chosen randomly in
$\theta_k\in[0,\pi)$.

\begin{figure}[b]
\centering
\includegraphics[angle=0,width=1.0\columnwidth]{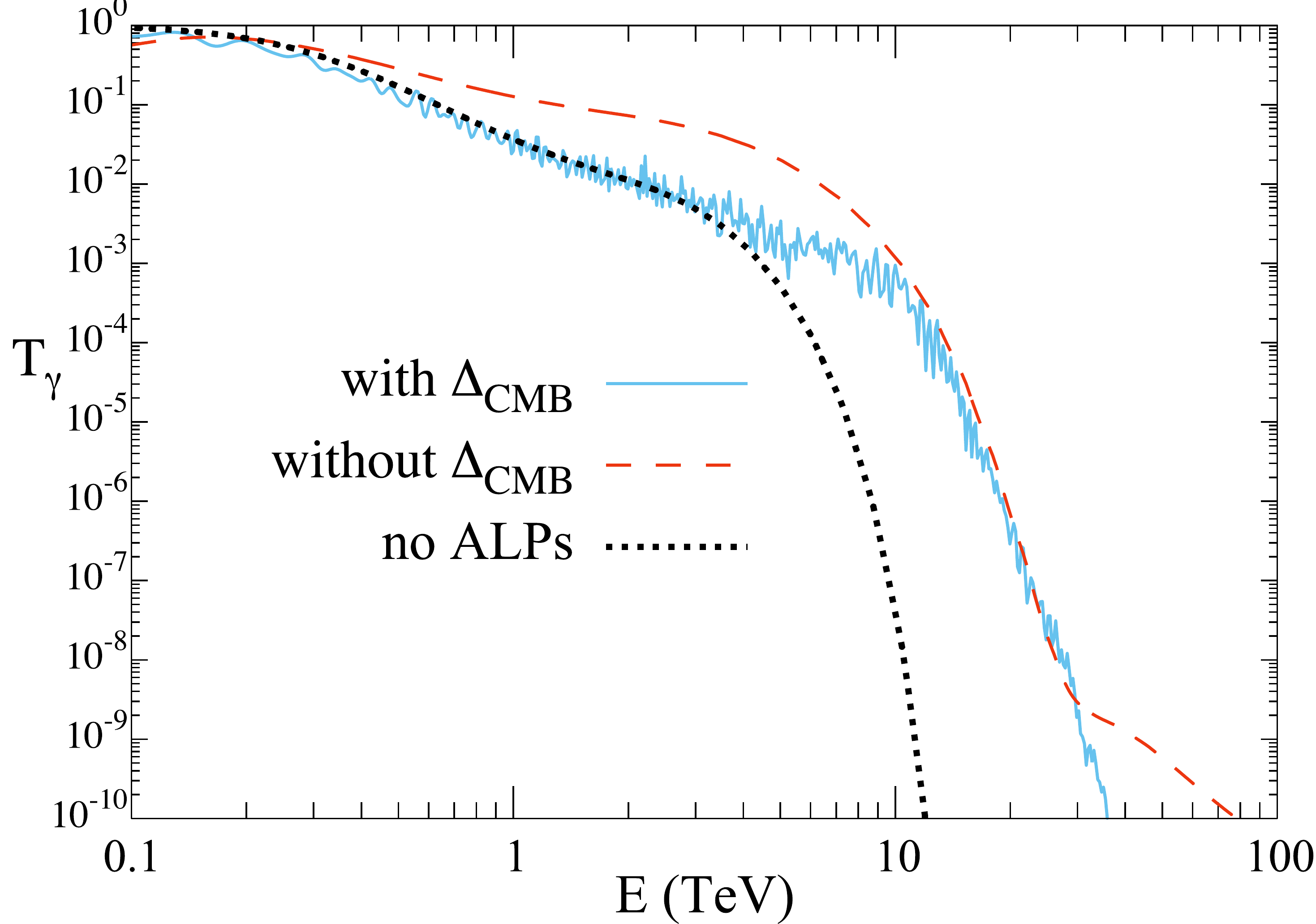}
\caption{The photon transfer function $T_\gamma$ as a
function of energy for a source at redshift $z=0.3$ for $m_a
=  10^{-10}$~eV and $\gag = 4 \times 10^{-12}$~GeV$^{-1}$
(LM) for a particular realization with (continuous blue
curve) and without (dashed red curve)  $\Delta_{\rm CMB}$. 
The dotted curve corresponds to the case of only absorption
onto EBL.}
\label{fig:CMBterm}
\end{figure}

\begin{figure*}[t]
\begin{centering}
\begin{minipage}[t]{1.\columnwidth}
\includegraphics[width=1.0\textwidth]{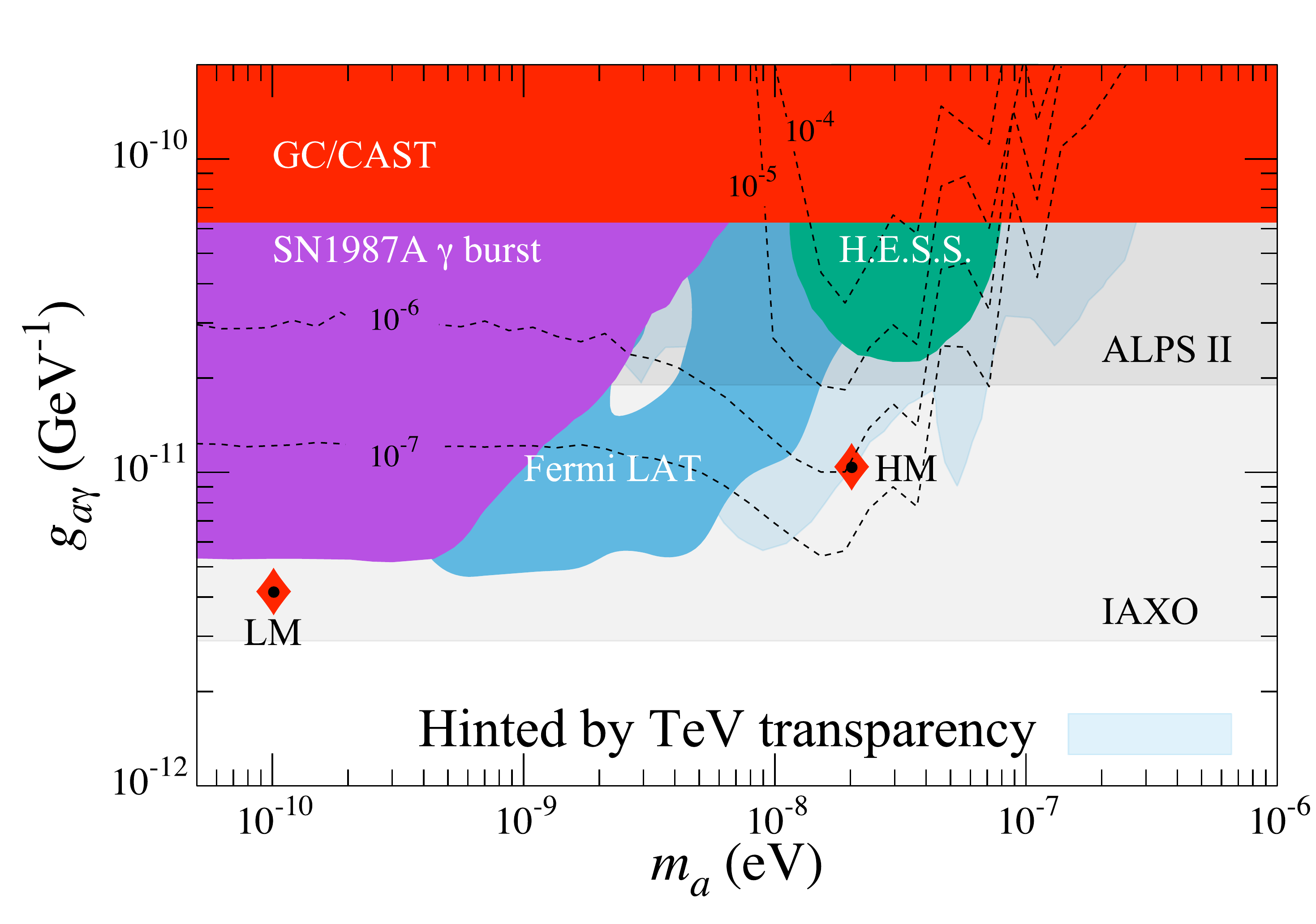}
\end{minipage}
\hfill
\begin{minipage}[t]{1.\columnwidth}
\includegraphics[width=1.0\textwidth]{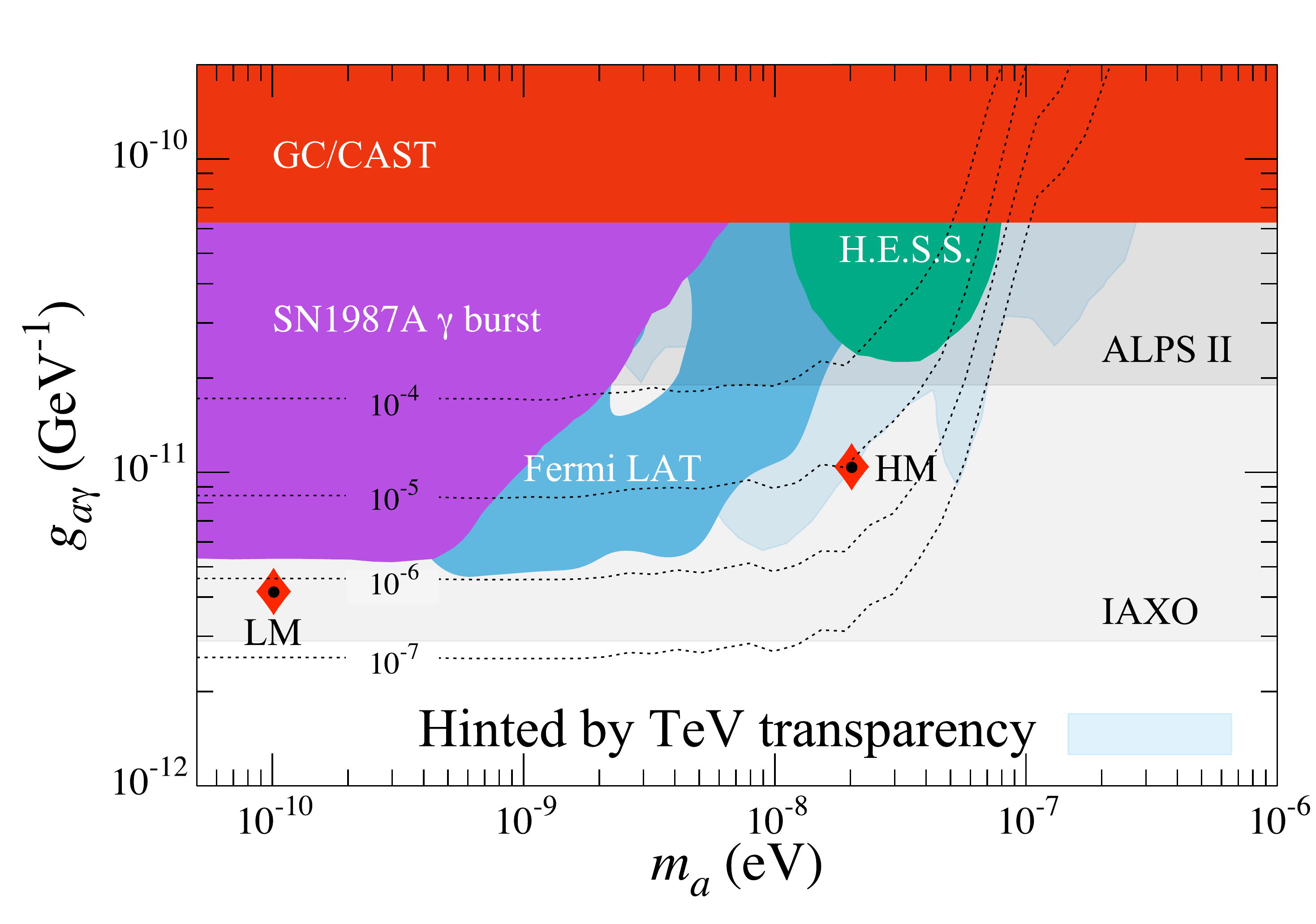}
\end{minipage}
\end{centering}
\vspace{-0.4cm}
\caption{Iso contour lines for transfer function 
$T_{\gamma} $ for a source at $z=0.3$ and at an energy
$E=20$~TeV for cell model (left) and realistic model (right)
of $B$-field. Se the text for more details. }
\label{fig:Iso}
\end{figure*}

Working in terms of the Eq. (\ref{vne}), after the
propagation over $n$ magnetic domains the density matrix is
given by repeated use of ${\cal H}_{\rm disp}^k$, namely
\begin{equation}
\label{mr1}
\rho_n = T ({\bf b}_n,\ldots , {\bf b}_1) \, \rho_0 \,
T^{\dagger} ({\bf b}_n, \ldots , {\bf b}_1)~,
\end{equation}
where we have set
\begin{equation}
T ({\bf b}_n, \ldots , {\bf b}_1)  \equiv \prod^n_{k = 1}
T_k~,
\end{equation}
 with
\begin{equation}
T_k = e^{(i {\cal H}_{\rm disp}  - \frac{i}{2} {\cal H}_{\rm
abs})  l_c} \,\ ,
\end{equation}
which is the transfer function in the $k$-th domain. In a
cosmological context however we should remember that $B_0$
and $l_c$ are no longer fixed but scale as $B_0\,
(1+z_k)^2$ and $l_c/ (1+z_k)$ where $z_k$ is the redshift of
the cell.

In the realistic case we solve the same equations, but the
field ${\bf B}$ is no longer a random vector but its three
components are calculated from the numerical model for
every realization. 
 
\emph{The effect of the $\Delta_{\rm CMB}$ term }---We 
briefly comment on the role of $\Delta_{\rm CMB}$ term on
the ALP-photon conversions at high energies. This term due 
to VHE photon  refraction on CMB photons been recently
calculated in~\cite{Dobrynina:2014qba}. From
Eqs.~(\ref{eq:deltaosc}) and (\ref{conv})  it results that
assuming $\Delta_{\rm CMB}=0$, when $\Delta_{a\gamma} \gg
\Delta_{a}, \Delta_{\rm pl}$  $P_{a\gamma}$ would become
energy-independent. Conversely, including  $\Delta_{\rm
CMB}$  this term can mimic a mass term, producing peculiar
energy-dependent features.  In order to illustrate this
effect, in Fig. \ref{fig:CMBterm} we show the transfer
function $T_\gamma$  as a function of energy for a source at
redshift $z=0.3$ for $m_a =  10^{-10}$~eV and $\gag = 4
\times 10^{-12}$~GeV$^{-1}$ (LM case) in presence of ALP
oscillations including the $\Delta_{\rm CMB}$ effect
(continuous curve) and without it (dashed curve). For
comparison it is also shown  $T_\gamma$  with only 
absorption on EBL (dotted curve). As predicted  $\Delta_{\rm
CMB}$ is responsible for the energy-dependent ``wiggles" in
the $T_\gamma(E)$ which are absent when $\Delta_{\rm
CMB}=0$. Another important consequence of $\Delta_{\rm CMB}$
is to suppress the transfer function at high energies when
$\Delta_{\rm CMB}\gtrsim \Delta_{a\gamma}$ (at $E > 30$~TeV
in Fig.~\ref{fig:CMBterm}).

\emph{Transfer function in the $(\gag,m_a)$ space}---In 
Fig.~\ref{fig:Iso} we show, superposed to the exclusion
regions of Fig.~1, curves iso-$T_{\gamma} $ for a source at
$z=0.3$ and at energy $E=20$~TeV. Left panel refers to the
cell model while right panel is for realistic magnetic
field. Each contour corresponds to the 95$^{\rm th}$
percentile of the distribution of $T_{\gamma} $. In other
words, there is a 5\%  probability that $T_{\gamma} $ is
larger than the indicated value. From these curves is
evident the enhancement of the area probed with the
realistic model at a fixed value of $T_{\gamma} $.

\newpage

\emph{Cosmological Simulations of Extragalactic
Magnetic Fields}---The simulation used in this work belongs to a dataset of
large cosmological simulations produced with the grid-MHD
code {\large E}{\small NZO} \cite{Enzo14}, presented in detail
\cite{Vazza14} and \cite{Gheller16} and designed to study
the evolution of extragalactic magnetic fields under
different physical scenarios.  This simulation employed
non-radiative physics to evolve a comoving volume of $200^3
\rm Mpc^3$, assuming a cosmology with $H=67.8 \rm~km/(s
\cdot Mpc)$, $\Omega_b=0.0478$, $\Omega_{\rm tot}=1.0$,
$\Omega_\Lambda=0.692$.  The magnetic field has been
initialized to the uniform value of $B_0=1 ~\rm nG$ along
each coordinate axis at the begin of the simulation
($z=38$).  With its $2400^3$  cells/dark matter particles
(for the fixed spatial resolution of $83.3$ kpc per cell)
this dataset represents the largest magnetohydrodynamical
simulation in the literature so far.
\begin{figure}[H]
\centering
\vspace{-2.8cm}
\includegraphics[angle=0,width=0.9\columnwidth]{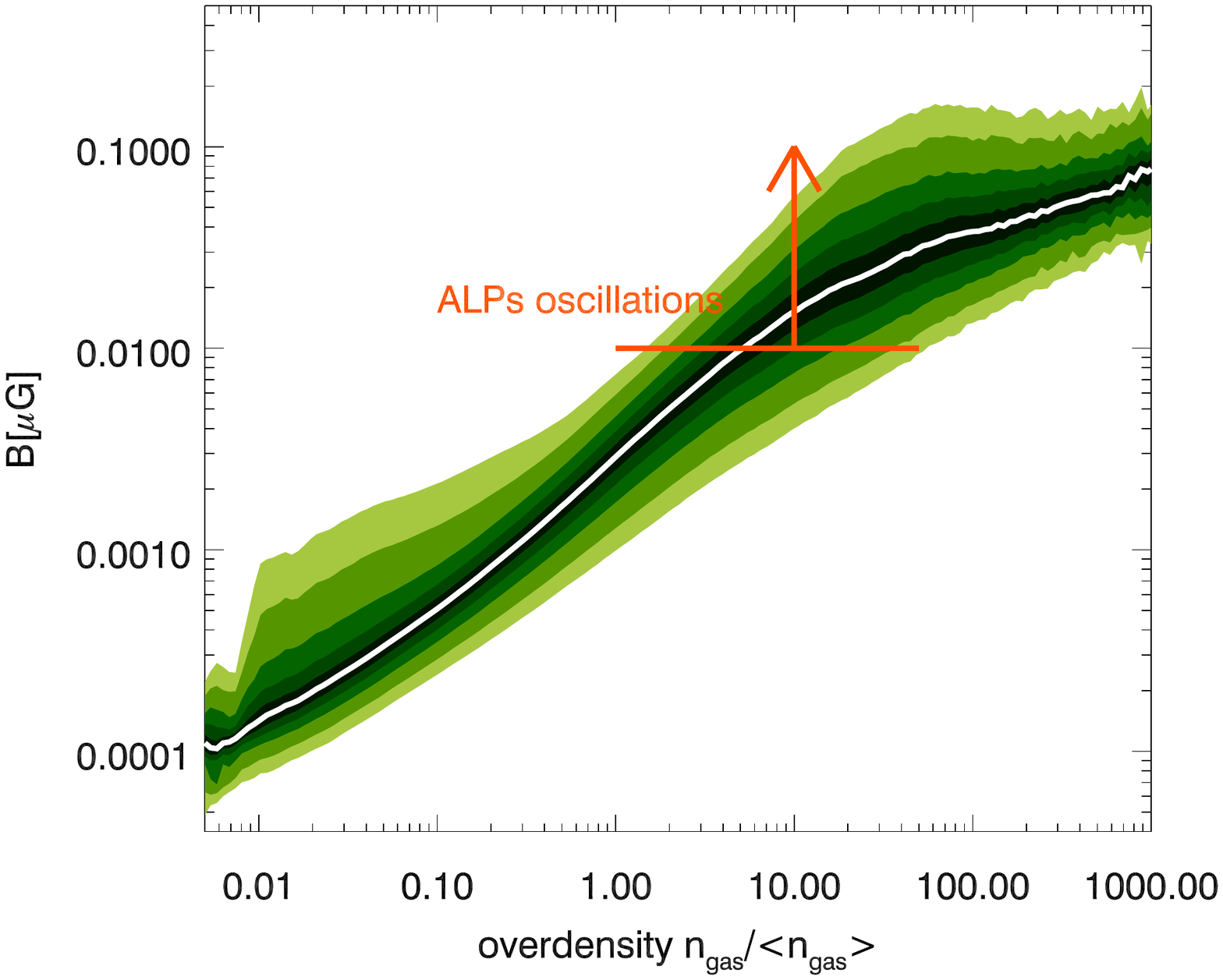}
\vspace{-2.8cm}
\caption{Distribution of typical magnetic field strength as
a function of gas overdensity for a representative samples
of line of sights through our simulated volume. The
different lines mark the percentiles of the distribution at
each overdensity, tenfold from 10 to 90\%. The additional
orange arrow approximately marks the regime in which we
observe significant photon-ALPs conversion in the energy
range investigated in the  paper.}
\label{fig:cosmo1}
\end{figure}

Fig.~\ref{fig:cosmo1} shows the distribution of magnetic
fields as a function of cosmic environment in our
simulation. The majority of the investigated volume presents
a magnetisztion level that follows the
compression/rarefaction of gas: $|{\bf B}| \propto (n_{\rm
gas}/\langle n_{\rm gas} \rangle)^{2/3}$, i.e. the
frozen-field approximation. The additional scatter on the
relation is due structure formation dynamics; in particular
the larger fluctuations at the high-density end of the
distribution are a product of small-scale dynamo
amplification acting within virialized halos.

\begin{figure}[H]
\centering
\vspace{-2.8cm}
\includegraphics[angle=0,width=0.9\columnwidth]{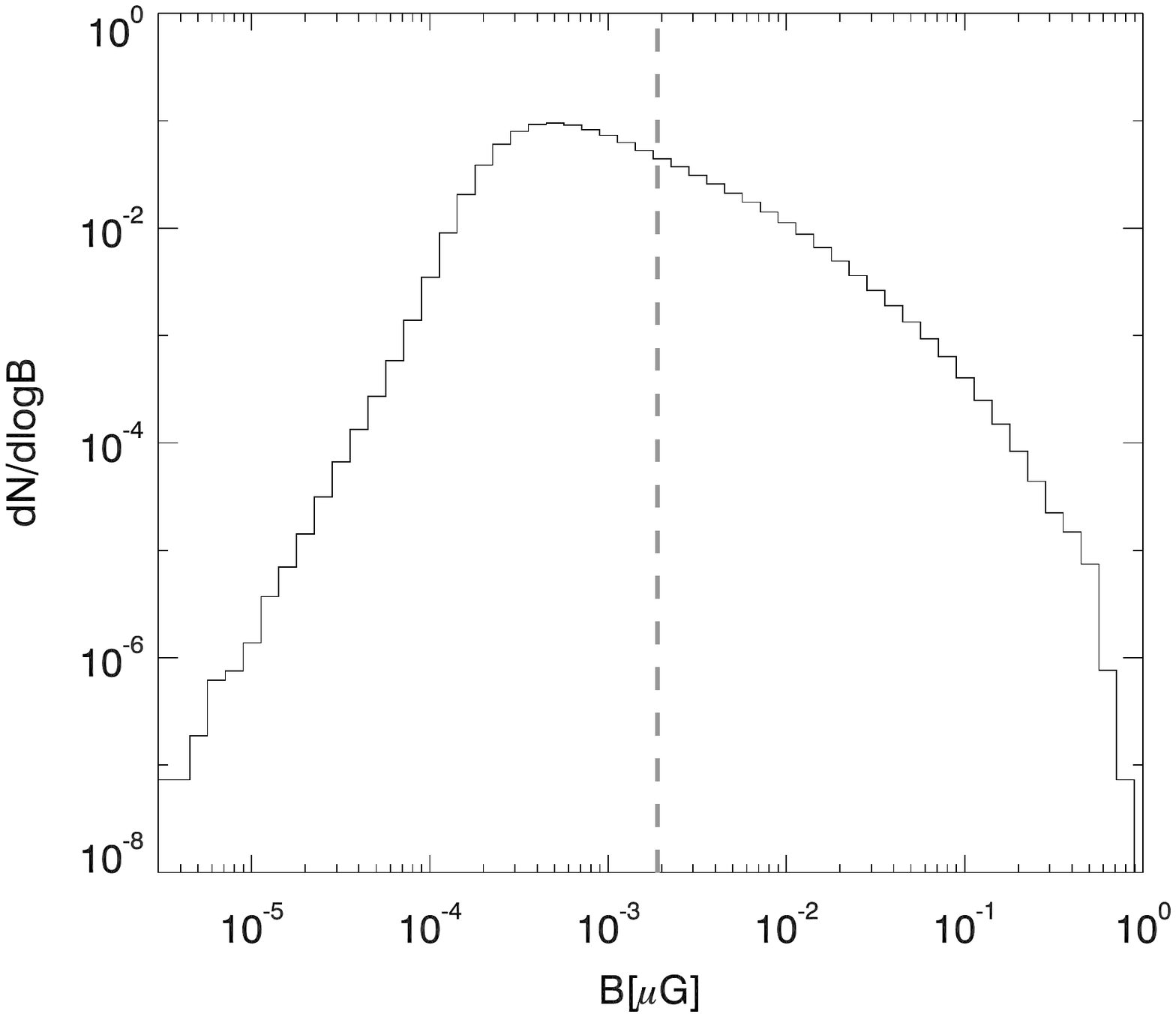}
\vspace{-2.8cm}
\caption{Volume distribution of typical magnetic field
strength for the same set of simulated lines of sight as in
Fig.~\ref{fig:cosmo1} (black). The additional dashed line
show the r.m.s.\ magnetic field value for the same
distribution.}
\label{fig:cosmo2}
\end{figure}

Fig.~\ref{fig:cosmo2} gives the volume distribution of
magnetic field strength for the same set of lines of sight.
The majority of the volume has a magnetization level
slightly below the initial seed field (as an effect of
adiabatic expansion in voids), yet a pronounced tail with 
magnetic fields up to $\sim \mu$G is present, largely
exceeding the r.m.s.\ magnetic field measured within the
volume, i.e. $1.9 \times 10^{-9} \rm G$ (as shown with the
vertical grey line).  The distribution of extragalactic
magnetic fields reproduced in this simulation is a first,
important step towards the simulation of possible scenarios
for the origin of cosmic magnetic fields, which is presently
limited by the constraints from the  cosmic microwave
background ~\cite{Planck2015}, and it will also possibly
include the magnetization from astrophysical sources, like
radio galaxies, starburst winds and jets from active
galactic nuclei. The impact of such mechanisms on the
conversion of ALPs will be subject of forthcoming work.


\end{document}